\documentclass[twocolumn,secnumarabic,amssymb, nofootinbib, aps, pra]{revtex4-2}
\usepackage{amsmath,amsthm}
\usepackage{url}

\newcommand{\bN}{{\mathbb N}}

\newcommand{\bp}{{\mathbf p}}

\newcommand{\Bf}{{\mathbf f}}
\newcommand{\Be}{{\mathbf e}}
\newcommand{\tbe}{\tilde{{\mathbf e}}}
\newcommand{\tq}{\hat{Q}}

\renewcommand{\S}{{\mathbf{S}}}
\renewcommand{\H}{{\mathcal{H}}}

\newcommand{\be}{\begin{equation}}
\newcommand{\ee}{\end{equation}}

\newcommand{\BC}{\mathbb{C}}

\def\f{{\bf f}}
\def\g{{\bf g}}
\newcommand{\bpi}{\boldsymbol{\pi}}
\def\bM{{\bf M}}
\def\bN{{\bf N}}

\begin{document}
\title{A Comment on the ``Photon position operator with commuting components'' by Margaret Hawton }
\author{A. Jadczyk}
 \email{ajadczyk@physics.org}
\affiliation{Laboratoire de Physique Th\'{e}orique, \\Universit\'{e} de Toulouse III \& CNRS\\
and\\
Ronin Institute, Montclair, NJ 0704}
\author{A. M. Schlichtinger}
\affiliation{Faculty of Physics and Astronomy, University of Wroclaw\\
 pl. M. Borna 9 50-204 Wroclaw, Poland}
\date{\today}
\begin{abstract}
We show that the position operator with commuting components proposed by M. Hawton [M. Hawton, Phys. Rev. A {\bf 59}, 954 (1999)] and developed in subsequent papers, including the recent ones,  does not have the properties required for a photon position operator. Depending on the exact interpretation of the results it either concerns a triplet of massless spin zero particles rather than the photon, or does not have the required covariance properties under space rotations.
\end{abstract}
\maketitle
\section{Introduction}
The problem of localizability of photons is usually discussed in term of existence of a self-adjoint position operator $\bf{Q}$ having the required natural commutation relations with the generators of space translations, space rotations as well as the space and time inversions within a unitary representation of the Poincar\'{e} group acting on the Hilbert space of the photon states. It is well known (cf. e.g. \cite{mourad} and references therein) that there is a unique solution to this problem: the so-called Pryce operator \cite{pryce} with noncommuting components. \footnote{The nonexistence of such a position operator with commuting components has been proven with a complete mathematical rigor (using only bounded operators: unitary operators and orthogonal projection operators)  in \cite[Theorem 2, p. 157]{amrein}.}

Notwithstanding these rigorous mathematical results, in several papers (cf. \cite{h1,h2,h3}), authored by M. Hawton et al., a statement has been made that there exists a whole family of self-adjoint position operators satisfying the requirements set in \cite{mourad,amrein}, yet having, nevertheless, commuting components. We have examined carefully this evident contradiction and have found that it has its source in insufficient mathematical and logical precision of the interpretation of the results obtained in Ref. \cite{h1}. It is our intent to fill this precision gap, and to prevent further propagation of the error made in \cite{h1} and employed recently in the recent studies of the photon Berry's phase problem   \cite{h3}and  of photon quantum mechanics in real Hilbert space \cite{h4}.\footnote{The error propagated also into PhD theses, seee.g. Ref. \cite{debierre}.}

In Sec. \ref{sec:la} we introduce the Lie algebra of the Poincar\'{e} group and its particular realization in the Hilbert space $\H$ of vector valued complex functions $\Bf(\bp)$ with Lorentz non-invariant measure $d^3p.$ We define the Pauli-Lubanski pseudovector $W^\mu$ and the helicity operator $\Lambda$, defined by $W^\mu=\Lambda P^\mu$, and we split $\H$ into photon subspace $\H_{ph}$ and its orthogonal complement $\H_0$ of spin zero states. In Sec. \ref{sec:po} we state the standard conditions for photon position operator and define the Pryce operator (with noncommuting components) - the unique solution of these conditions. In Sec. \ref{sec:pb} we define the polarisation basis and introduce the position operator as proposed in Ref. \cite{h1}. In Sec. \ref{sec:ps} we identify the problem with the the false staements made in Ref. \cite{h1}. Since \cite{h1} lacks clarity concerning the selection of the particular representation of the Poincar\'{e} group, we analyze all thee available representations and show that the position operator defined in \cite{h1} does not have the required properties in all three cases (though in each case for a different reason).
\subsection{Notation}We use the Minkowski metric $(\eta_{\mu\nu})=(\eta^{\mu\nu})=\mbox{diag}(+1,-1,-1,-1)$. We will use this metric to rise and lower indices of vectors and tensors. Our $3$-vectors are, as a rule, contravariant vectors with upper indices. Thus, for example, ${\bf p}$ is a vector with components $(p^1,p^2,p^3)$ etc. The three spin matrices $\S$ are always written with lower indices $\S=(S_1,S_2,S_3).$ To simplify the notation we use a system of units in which $c=1$ and $\hbar=1.$
\section{Photon wave function\label{sec:pwf}}
Photons are defined through a particular unitary representation of the Poincar\'{e} group characterized by mass zero and helicity $\pm 1.$ It is, however, convenient to include states with helicity zero as well.
\subsection{The Lie algebra\label{sec:la}}
In any unitary representation the ten self-adjoint generators of the Poincar\'{e} group satisfy the following commutation relations:
\begin{gather}\nonumber [P_\mu,P_\nu]=0,\, [P_\mu,M_{\rho\sigma}]=i(\eta_{\mu\rho}P_\sigma-\eta_{\mu\sigma}P_\rho),\\ [M_{\mu\nu},M_{\rho\sigma}]=i(\eta_{\nu\rho}M_{\mu\sigma}-\eta_{\mu\rho}M_{\nu\sigma}+
\eta_{\mu\sigma}M_{\nu\rho}-\eta_{\nu\sigma}M_{\mu\rho}).\notag\end{gather}
The rotation and the boost generators ${\bf M}$ and ${\bf N}$ are then defined as
$ M^i=\frac12 \epsilon^{ijk}M_{jk},\, N^i=M^{0i}=M_{i0}.$
Thus, for instance, $M^1=M^{23}=M_{23}$, etc.

Let ${\cal H}$ be the Hilbert space of square integrable
$3$--component complex functions on $\mathbb{R}^3$ with
a scalar product\footnote{
In Ref. \cite{h1} a family of scalar products with $d^3p/p_0$ replaced by $d^3p/p_0^{2\alpha}$ is being considered. They all lead to unitarily equivalent representations of the Poincar\'{e} group. The standard Lorentz invariant scalar product is obtained  by setting $\alpha=\frac12.$ Here we choose $\alpha=0.$
} \be (\f ,\g )=\sum_{i=1}^3 \int f_i^\star(\bp)g_i
(\bp)d^3p.\ee

With the notation $\bp =(p^1,p^2,p^3),$ $p^0=p_0=\left(
\sum_{i=1}^3 p^i\right)^{\frac{1}{2}},$
$\bpi = \bp /p^0,\ \  {\bf S} = (S_1,S_2,S_3),$
$$S_1=\left(\begin{smallmatrix}0&0&0\\0&0&-i\\ 0&i&0\end{smallmatrix}\right),\,
S_2=\left(\begin{smallmatrix}0&0&i\\ 0&0&0\\ -i&0&0\end{smallmatrix}\right),\,
S_3=\left(\begin{smallmatrix}0&-i&0\\ i&0&0\\ 0&0&0\end{smallmatrix}\right)$$
the generators $P^\mu, $ $M^i=\frac{1}{2}\epsilon_{ijk}M^{jk}, and $
$N^i=M^{0i}$ of the Poincar\'e group are given by
\be P^\mu=p^\mu,\,
\bf{M}=\bf{L}+\bf{S},\, \bN=\bf{K}+\bf{n},\label{eq:M}\ee
where
\be {\bf L}= -i(\bp\times
\partial /\partial \bp )\label{eq:L},\,
{\bf K}= i(p_0\, \partial /\partial \bp+\frac12\bpi),\ee
 \be
 \bf{n}=\bpi
\times \bf{S} .\label{eq:n}\ee
The above operators satisfy the following commutation relations:
\be\nonumber [N^i,p^j]=i\delta^{ij}p^0,\,
 [N^i,p^0]=ip^i,\,
[M^i,p^j]=i\epsilon^{ijk}p^k,\ee
\be\nonumber [M^i,p^0]=0,\, [N^i,N^j]=-i\epsilon^{ijk}M^k,\ee
\be\nonumber [M^i,N^j]=i\epsilon^{ijk}N^k,\,[M^i,M^j]=i\epsilon^{ijk}M^k.\ee
Replacing $\bf{M}$ by $\bf{L}$ and $\bf{N}$ by $\bf{K}$ we obtain the same commutation relations.

The unitary space inversion operator $\Pi$ and antiunitary time inversion operator $\Theta$ for this representation  are given by (cf. \cite[Eq. (2.3)]{mourad}
$ (\Pi{\bf f})(\bp)={\bf f}(-\bp),$ $(\Theta{\bf f})(\bp)={\bf f}^*(-\bp).$
For a general representation of the Poincar\'{e} group one defines the four-dimensional Pauli-Lubanski pseudovector $W^\mu$ as
$ W_\mu=\frac12 \epsilon_{\nu\rho\sigma\mu}\,P^\nu M^{\rho\sigma}.$
We have
$ W^0={\bf P}\cdot{\bf M},\, {\bf W}=P^0\,{\bf M}-{\bf P}\times {\bf N}.$

It follows from the very definition that $\eta_{\mu\nu}P^\mu W^\nu=0.$  For mass zero representations $P^\mu$ is lightlike, and therefore $W^\mu$ is proportional to $P^\mu$:
$ W^\mu=\Lambda P^\mu.$
The proportionality operator $\Lambda$ commutes with all the generators and is called the helicity operator (see e.g. \cite[p.64]{ryder}).
In our case one easily finds that for generators $P^\mu,\bf{L},\bf{K}$ we have $W^\mu=0$, therefore $\Lambda=0,$ while for the generators $P^\mu,\bf{M},\bf{N}$ we have
\be \Lambda=\bpi \cdot\bM=\bpi \cdot\bf{S}.\label{eq:lambda}\ee
From the explicit form $(\Lambda \f
)=-i\bpi\times \f $ it is easy to see that the spectrum of
$\Lambda$ is discrete and consists of three points $\lambda=\pm 1$
and $0.$ Therefore $\Lambda^2$ is a projection onto the subspace $\H_{ph}$ of $\H.$ $\H_{ph}$ is a direct sum of eigenspaces
$\H_{\pm }$ of $\Lambda$ corresponding to eigenvalues
$\lambda=\pm
1.$ The photon states are represented by vectors in $\H_{ph}.$ The
orthogonal complement $\H_0$ of $\H_{ph}$ in $\H$ describes a
spinless particle. It easily follows from these definitions that

$
\H_{ph}=\{\f\in\H:\,\bp\cdot\f (\bp )=0\},$
and that
$\H_0=\{\f\in\H:\,\f (\bp )=c(\bp)\bp\}$
for some scalar function $c(\bp).$\\ It is obvious from the above
that $P_0$ and $M_{\mu \nu }$ leave the subspaces $\H_{ph}$ and
$\H_0$ invariant.
\subsection{The Pryce position operator\label{sec:po}}
The standard requirements for the selfadjoint photon position operator $X^i=X^{i*}$ involve covariance under space translations and rotations as well as invariance under space and time reflections, and that $X^i$ leave the photon subspace $\H_{ph}$ invariant.  They are expressed by the following formulas:
\be \begin{split}
[M^i,X^j]&=i\epsilon^{ijk}X^k,\, [P^i,X^j]=-i\delta^{ij},\\
\Pi X^i\Pi&=-X^i,\,\Theta X^i\Theta =X^i,\,
[X^i,\Lambda^2]=0.
\end{split}\label{eq:pp}\ee
The last condition is equivalent to: if $\f(\bp)=c(\bp)\,\bp,$ where $c(\bp)$ is any scalar valued complex function of $\bp$, then $X^i\f(\bp )=c^i(\bp)\,\bp,$ for some functions $c^i(\bp),$ which is easier to verify in practice.

It is shown in Ref. \cite{mourad} that the above problem has a unique solution. It is the so-called Pryce operator ${\bf X}_P$ given by
\be X_P^i=i\frac{\partial}{\partial p^i}+\sum_{j,k}\epsilon_{ijk}\frac{P^jS_k}{P_0^2}.\ee
\subsection{Polarization basis\label{sec:pb}}
Let $\Be_i(\bp)$ the triplet of vectors tangent to the coordinate lines $p_1,p_2,p_3$
\be\Be_1=\left(\begin{smallmatrix}1\\0\\0\end{smallmatrix}\right),\,
\Be_2=\left(\begin{smallmatrix}0\\1\\0\end{smallmatrix}\right),\,
\Be_3=\left(\begin{smallmatrix}0\\0\\1\end{smallmatrix}\right).\ee
The procedure of constructing the position operator with commuting components, as proposed in Ref. \cite{h1} starts with a choice of another $\bp$-dependent (orthonormal in $\BC^3$) basis $\tbe_i(\bp)$, with
$\tbe_1(\bp),\tbe_2(\bp)$ almost everywhere differentiable functions of $\bp$ and orthogonal,  while $\bp$ and $\tbe_3(\bp)=\bp/|\bp|$. Apart of these requirements the basis needs not be unspecified.
\footnote{In Refs. \cite{h1,h2,h3} the basis is a real basis, and specified as related to polar coordinates in the momentum space - see below. But the essence of the construction of the `position operator with commuting components' does not depend on any particular choice of the basis, except of the requirements mentioned above.}
\footnote{Since the (complexified) tangent bundle of the sphere $S^2$ is nontrivial (see e.g. \cite{chiao}), the functions $\tbe_i(\bp)$ are defined only almost everywhere. Notice that they are `improper' (unnormalizable) elements of $\H .$}

Let $U(\bp)$ be the unitary operator in $\BC^3$ defined by \be
U(\bp)\Be_i(\bp)= \tbe_i(\bp),\,
U^*(\bp)\tbe_i(\bp)= \Be_i(\bp)\quad(i=1,2,3)\label{eq:uee}\ee
for almost all $\bp$. Then $U^*(\bp)\tbe_i(\bp)=\Be_i(\bp)$ and the matrix-valued function $U(\bp)$ defines a unitary operator $U$ in $\H.$
The operator $U$ maps the subspaces generated by $\Be_i$ onto the subspaces generated by $\tbe_i,$ while $U^*$ does the converse.

Let the hermitian operators $Q^i$ on $\H$ be defined as
\be Q^i=i\frac{\partial}{\partial p^i}.\label{eq:Q}\ee
The operators $Q^i$ evidently commute and leave the subspaces generated by $\Be_i$ invariant. But they do not leave the subspace $\H_{ph}$ invariant, what constitutes the main problem with photon localization.

We notice at this point that the operators $K^i$ defined by Eq. (\ref{eq:L}) can be written as
\be K^i=\frac12 (Q^iP^0+P^0 Q^i).\label{eq:KQ}\ee
Following Ref. \cite{h1} (cf. also \cite{h3}) we introduce the operators $\tq^i$ defined as
\be \tq^i=UQ^iU^*.\label{eq:hqi}\ee
Then $\tq^i$ are self-adjoint, commute and leave the subspaces generated by $\tbe_i$ invariant. In particular $\tq_i$ leave the subspace $\H_{ph}$ invariant.\footnote{They are denoted $\mathbf{r}_i$ in \cite{h1} and by $\mathbf{x}_i$ in \cite{h3}.}

Hawton \cite{h1,h2,h3} specifies the polarization basis to be the one induced by spherical coordinate system in momentum space, with
$$
p^1=|\bp|\sin\theta\cos \phi,\,
p^2=|\bp|\sin\theta\sin\phi,\,
p^3=|\bp|\cos\theta.
$$
Let $\bpi=\bp/|\bp|,\hat{\boldsymbol{\theta}}, \hat{\boldsymbol{\phi}}$ be the unit tangent vectors to the coordinates of the spherical coordinate system. Following Ref. \cite{h1}
 we call them $\tbe_3,\tbe_1,\tbe_2.$ Explicite:
 \begin{gather}
\tbe_1(\bp)=\left(\begin{smallmatrix}\cos\theta\cos\phi\\ \cos\theta\sin\phi\\-\sin\theta\end{smallmatrix}\right),\,
\tbe_2(\bp)=\left(\begin{smallmatrix}-\sin\phi\\ \cos\phi\\0\end{smallmatrix}\right),\notag\\
\tbe_3(\bp)=\bp/|\bp|=\left(\begin{smallmatrix}\sin\theta\cos\phi\\ \sin\theta\sin\phi\\ \cos\theta\end{smallmatrix}\right).\end{gather}
We then define the matrix $U$ whose columns are these vectors:
\be U=\begin{pmatrix}\cos\theta\cos\phi&-\sin\phi&\sin\theta\cos\phi\\
\cos\theta\sin\phi&\cos\phi&\sin\theta\sin\phi\\
-\sin\theta&0&\cos\theta\end{pmatrix}.\ee
Then $U$ satisfies Eqs. (\ref{eq:uee}). The matrix $U$ is real, therefore $U^*=U^t.$
We notice that the third column describes a spinless particle. A straightforward calculation, using Eq. (\ref{eq:hqi}), leads to the explicit form of $\hat{\mathbf Q}$:
\be \hat{\mathbf Q}=i\frac{\partial}{\partial \bp}+\frac{1}{p_0^2}\bp\times {\mathbf S}-\frac{\cot \theta}{p_0}\tbe_2 S_3,\ee
in agreement with Eq. (1) in \cite{h1}.
\section{The problem and its solution\label{sec:ps}}
Following the definitions and the notation in Ref. \cite{h1} we introduce a new representation of the Poincar\'{e} group defined by self-adjoint generators $\hat{P}^\mu,\, \hat{L}^i,\, \hat{K}^i$  given by:
\be
\hat{P}^\mu=UP^\mu U^*,\,
\hat{L}^i=UL^i U^*,\,
\hat{K}^i=UK^i U^*.
\ee
First of all we notice that, since $U$ is a multiplication by a $\bp$-dependent matrix, it commutes with $P^\mu$. Therefore we have that $\hat{P}^\mu=P^\mu.$ Now, since $\bf{L} =\bf{Q}\times\bf{P}$ (cf. Eq (\ref{eq:L})), we have
\be \hat{\bf{L}}=\hat{\bf{Q}}\times \hat{\bf{P}},\label{eq:hl}\ee
in agreement with Eq (17) of Ref. \cite{h1}. Similarly, from Eq. (\ref{eq:KQ}), we obtain
\be \hat{\bf{K}}=\frac12 (\tq P^0+P^0\tq)\label{eq:hk}\ee
in agreement with Eq. (18) in \cite{h1}.

The unitary representation of the Poincar\'{e} group defined by the generators $\hat{P}^\mu,\, \hat{L}^i,\, \hat{K}^i$ describes a triplet of particles of mass zero (since $\hat{P}^2=P^2=0$) and spin zero (since the Pauli-Lubanski pseudovector $\hat{W}_\mu,$ calculated for this representation, is identically zero). I does nor represent photon's boosts and rotations.

There is also another unitary representation determined by generators
\be \tilde{P}^\mu=\hat{P}^\mu=P^\mu,\, \tilde{{\bf M}}=U{\bf M}U^*,\,\tilde{{\bf N}}=U{\bf N}U^*,\label{eq:tl}\ee
unitarily equivalent to $P^\mu,\bf{M},\bf{N},$ and describing photons of both helicities and a scalar spin zero particle. From Eqs. (\ref{eq:M})-(\ref{eq:n}) we have\footnote{The operators $\tilde{S}_i$ are denoted $S_{\bp i}$ in \cite{h1}.}
$\tilde{\bf{M}}=\hat{\bf{M}}+\tilde{\bf{S}},$
where
$\tilde{\bf{S}}=U{\bf{S}}U^*.$
The helicity operator $\tilde{\Lambda}$ for this representation is given by
$\tilde{\Lambda}=U\Lambda U^*=\boldsymbol{\pi}\cdot\tilde{\bf{S}}.$
Its eigensubspaces are spanned by the vectors $U\tbe_i=U^2 \boldsymbol{\bf{e}}_{i}.$

It is stated explicitly in Ref. \cite{h1} that the operator $\hat{\bf Q}$ defined  by Eq. (\ref{eq:Q}) satisfies the conditions placed on a position operator in \cite{mourad,pryce,wigner, wightman}. Since this statement is in evident contradiction with the results of the cited references, we will now point out precisely where is the error in Ref. \cite{h1} and in the subsequent papers.

Since the conditions on the photon position operator mentioned above and written explicitly in Eqs. (\ref{eq:pp})
are all group-theoretical, it is necessary to list the relevant representations of the Poncar\'{e} group that are being under consideration. We have three such representations. While the translation generators $P^\mu$ are the same in all three cases, the generators of the homogeneous Lorentz group are different. We have the original representation with generators $M^i,N^i$, the representation with ``spinless'' generators $\hat{\bf L},\hat{\bf K}$ given by Eqs. (\ref{eq:hl}),(\ref{eq:hk}), and the complete transformed representation with generators $\tilde{\bf M},\tilde{\bf N}$ given by Eq. (\ref{eq:tl}).

Ref. \cite{h1} is not explicit about which representation of the Poincar\'{e} group is to be taken, therefore, in order to verify the conditions Eq. (\ref{eq:pp}), we will have to examine all three representations, and show which of these conditions is not satisfied by Hawton's position operator in each of the three cases.
\paragraph{The original representation ${\bf M},{\bf N}.$}
In this case the operator $\hat{\bf Q}$ is not covariant under rotations, $[M^i,\hat{Q}^j]\neq i\epsilon^{ijk}\hat{Q}^j.$ Indeed,  from Eq. (\ref{eq:pp}) we should have, in particular, $[M^1,X^1]=0.$ But the explicit calculation (using spherical coordinates in the momentum space) of this commutator on the vector $f(p_0,\theta,\phi)=(a(p_0),0,0),$ where $a$ is an arbitrary function of $p_0$ is evidently $\neq0.$
Thus Hawton position operator does not satisfy the condition of covariance under rotations of the original representation.
\paragraph{The representation $\hat{\bf M},\hat{\bf N}.$}
Here all the conditions in Eq. \ref{eq:pp}) are satisfied, but the Pauli-Lubanski pseudovector calculated for this representations is identically zero, thus the helicity operator {\em for this representation} is identically zero. The representation describes a triplet of spinless particles, not the photon.
\paragraph{The representation $\tilde{\bf M},\tilde{\bf N}.$}
In this case $\hat{\bf Q}$ is covariant under translations and rotations, but $\hat{\bf Q}$ {\em does not commute with the helicity operator $\tilde{\Lambda}$ of this representation\,} for the same reason for which ${\bf Q}$ does not commute with $\Lambda$ defined by Eq. (\ref{eq:lambda}). In fact, we have $\hat{\bf Q}=U{\bf Q}U^*,$ $\tilde{\Lambda}=U\Lambda U^*,$ therefore
\be [\hat{\bf Q},\tilde{\Lambda}^2]=U[{\bf Q},\Lambda^2]U^*.\ee
Since $[{\bf Q},\Lambda^2]\neq 0,$ we also have $[\hat{\bf Q},\tilde{\Lambda}^2]\neq 0.$

\end{document}